\documentclass[aps,prl,reprint,superscriptaddress,showpacs]{revtex4-1}
\usepackage[T1]{fontenc}
\usepackage[latin9]{inputenc}
\usepackage{amsmath}
\usepackage{amssymb}
\usepackage{graphicx}
\usepackage{esint}
\usepackage{bm}
\usepackage{amsfonts,color}
\usepackage{epstopdf} 
\usepackage{hyperref}

\begin{document}

\title{
Reduced Low Dose Rate Sensitivity and its Mechanism in Bipolar Junction Transistors
}

\author{Xiao-Long Feng$^\dag$}
\affiliation{Microsystem and Terahertz Research Center, China Academy of Engineering Physics,
Chengdu 610200, P.R. China}
\affiliation{Institute of Electronic Engineering, China Academy of Engineering Physics,
Mianyang 621999, P.R. China}

\author{Zi-Bo Wang$^\dag$}
\affiliation{Microsystem and Terahertz Research Center, China Academy of Engineering Physics,
Chengdu 610200, P.R. China}
\affiliation{Institute of Electronic Engineering, China Academy of Engineering Physics,
Mianyang 621999, P.R. China}

\author{Yang Liu$^\dag$}
\affiliation{Microsystem and Terahertz Research Center, China Academy of Engineering Physics,
Chengdu 610200, P.R. China}
\affiliation{Institute of Electronic Engineering, China Academy of Engineering Physics,
Mianyang 621999, P.R. China}

\author{Yu Song}
\email{songyu@mtrc.ac.cn}
\email{kwungyusung@gmail.com}
\affiliation{Microsystem and Terahertz Research Center, China Academy of Engineering Physics,
Chengdu 610200, P.R. China}
\affiliation{Institute of Electronic Engineering, China Academy of Engineering Physics,
Mianyang 621999, P.R. China}

\begin{abstract}
It is surprising that only an enhanced low-dose-rate sensitivity (ELDRS), but not its contrariety, a reduced low-dose-rate sensitivity (RLDRS), is experimentally observed in bipolar junction transistors. In this work, we attribute this strong asymmetry to an overwhelming superiority of hydrogen cracking reactions relative to Shockley-Read-Hall recombinations. We demonstrate that this situation can be completely reversed and an RLDRS effect can occur by decreasing the concentration ratio between hydrogen and recombination centers. We show that the decrease of recombination rate of holes (generation rate of protons) with increasing dose rate is responsible for the positive (negative) dose-rate dependence of RLDRS (ELDRS). We also find an RLDRS-ELDRS transition for decreasing dose rate, in which the transition point can be controlled by the concentration ratio. The proposed RLDRS effect and its tunability pave a way for a credible acceleration test of the total ionizing dose damage in different low dose rate environments.
\end{abstract}

\date{\today}
\maketitle

\emph{Introduction.}--
Total ionizing dose (TID) effect is a phenomenon that
semiconductor devices degrade
when they are exposed to persistent ionizing irradiations.
In 1991, it was found that the degradation in some bipolar
junction transistors (BJT) becomes larger at a
lower dose rate for a given total dose \cite{Enlow1991Response}.
This phenomenon is termed as the enhanced low dose rate sensitivity (ELDRS) effect.
Due to this effect,
experiments should be carried out under a true dose rate,
which requires a rather long time for a relatively low dose rate
\cite{chen2010effects}.
For instance, it takes about one year to accomplish
a typical experiment
of a total dose of 30 krad and a dose rate of 1 mrad/s.

In efforts to solve this problem, several acceleration testing methods have been
proposed. For instance, people have tried to
elevate temperature during the irradiation \cite{fleetwood1994physical,johnston1996enhanced,pease1997proposed,
carriere2000evaluation,boch2004elevated},
alternate high dose rate irradiation and elevated temperature anneals
\cite{freitag1998study,pershenkov2007interface},
switch dose rate through experiments \cite{boch2004effect,boch2005estimation,boch2009use},
and, most recently, apply hydrogen externally
\cite{fleetwood2008electron,pease2008effects,adell2009irradiation}.
Although those proposals seem useful, 
limitations still exist in the real applications,
such as the lack of universal variables and so on \cite{adell2014dose}.

On the other hand, great efforts have also been made to reveal the underlying mechanism of
ELDRS.
In latest works, the process of TID is modeled as 
transports and reactions of electron-hole pairs induced by irradiation
and hydrogen 
mediated by shallow and deep oxygen vacancies \cite{adell2014dose}.
The defects can play roles of carrier recombination center
(Shockley-Read-Hall (SRH) recombination) and proton
release center (directly or through hydrogen cracking).
It is widely believed that the enhancement of the recombination rate with increasing dose rate
leads to a reduced high dose rate sensitivity (RHDRS),
which is equal to ELDRS \cite{hjalmarson2003mechanisms,boch2006dose,boch2006physical,
fleetwood2008electron,Hjalmarson2009Calculations}.
However, it is widely accepted that the SRH recombination rate decreases with
increasing difference of quasi-Fermi energies of electrons and holes \cite{shockley1952statistics,hall1952electron},
which is proportional to the dose rate.
The above mechanism also implies that the TID process is dominated by recombination reactions.

In this work, we demonstrate a different mechanism and
propose a contrariety of ELDRS, i.e., a reduced low dose rate sensitivity (RLDRS) effect.
We show that for conventional BJTs the concentration of H$_2$
is much higher than that of electrons induced by irradiation.
As a result, the hydrogen cracking reactions (HCR) dominant and
an ELDRS occurs because the generation rate of protons decreases with increasing dose rate.
The latter is a direct result of the Le Chatelier's principle.
We then demonstrate that, 
by decreasing the concentration ratio of hydrogen 
and recombination centers, the electron recombination reactions (ERR) can dominate.
Accordingly, an RLDRS effect arises through a decrease of the SRH recombination rate with increasing dose rate.
Taking advantage of the RLDRS effect, the TID damage of BJT at a low dose rate can be evaluated
with a relatively high dose rate with a dramatically shortened time cost.
We also find 
an RLDRS-ELDRS transition for decreasing dose rate.
The transition dose rate
can be controlled by the concentration ratio, 
which paves a way
for applications in different dose rate environments.
We indicate that, the RLDRS effect may have already been observed in an early experiment.

\emph{Model and implement.}--
In order to investigate the dose rate dependence of an interface trap density ($N_\textmd{it}$), which stands for the
damage of a BJT,
we adopt the ``quantitative'' model developed by Rowsey et al. 
\cite{rowsey2011quantitative,rowsey2012quantitative,rowsey2012mechanisms}.
In this model, shallow ($\delta$) and deep ($\gamma$) oxygen vacancies and their hydrogen complex
($V_{o\delta(\gamma)},V_{o\delta(\gamma)}H,V_{o\delta(\gamma)}H_2$) in the oxide are considered,
which interact with radiation induced carriers and impurities.
With the help of the finite 
element method, we implement this model in a python language \cite{TIDES}
and verify it by comparing the numerical results
with the experiments \cite{chen2007mechanisms,pease2008effects}.
We consider a base oxide of total thickness 1200nm and interface thickness 4nm
and a total dose of 30krd.
Using initial concentrations of different defects listed in Table.
~\ref{tab:initial0} and hydrogen-passivated dangling bond (Si-H) 
of $10^{13}$ cm$^{-2}$,
the calculation results match the experimental results well, including
$N_{it}$ vs. $H_2$ concentration \cite{chen2007mechanisms}
and $N_{it}$ vs. dose rate \cite{pease2008effects}. 

\begin{figure}[!t]
  \centering
  \includegraphics[width=0.9\linewidth]{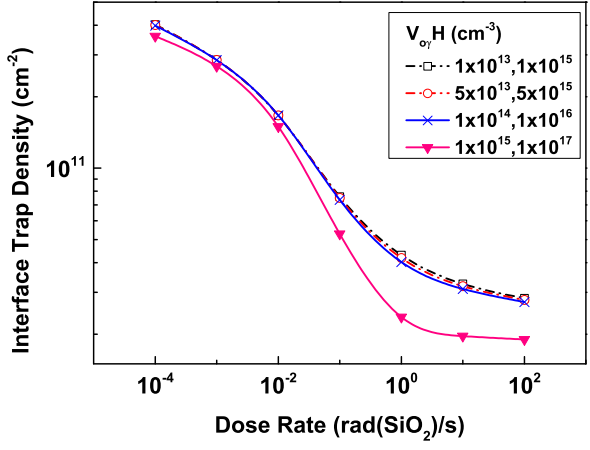}\\
  \caption{Dose rate dependence of $N_{it}$ for various concentrations of $V_{o\gamma}H$.
  The concentration of hydrogen is $10^{13}$cm$^{-3}$.
  }\label{fig:VogH0}
\end{figure}

\begin{table}[!h]
\caption{Initial concentrations of different defects and roles played by them.}
\label{tab:initial0}
\centering
\begin{tabular}{lccc}
\hline
Defects & Bulk & Interface & Roles \\
\hline
$V_{o\gamma}$ 
& $10^{14}$ cm$^{-3}$  & $9\times10^{18}$ cm$^{-3}$ & release center \\
$V_{o\gamma}H$ 
& $10^{14}$ cm$^{-3}$  & $10^{16}$ cm$^{-3}$ & recombination center \\
$V_{o\gamma}H_{2}$ 
& $10^{14}$ cm$^{-3}$  & $10^{18}$ cm$^{-3}$ & release center \\
$V_{o\delta}$ 
& $10^{18}$ cm$^{-3}$ & $10^{18}$ cm$^{-3}$ & recombination center \\
$V_{o\delta}H$ 
& $10^{14}$ cm$^{-3}$ & $10^{14}$ cm$^{-3}$ & release center \\
$V_{o\delta}H_{2}$ 
& $10^{15}$ cm$^{-3}$ & $10^{15}$ cm$^{-3}$ & release center \\
\hline
\end{tabular}
\end{table}

\emph{ELDRS with high $H_2$.}--
We first consider the dose-rate dependence of $N_{it}$
for [H$_2]=10^{13}$cm$^{-3}$ at various concentrations of $V_{o\gamma}H$ ($[V_{o\gamma}H]$), which is shown in Fig.~\ref{fig:VogH0}.
One can see that, for a given dose rate the value of $N_{it}$ decreases with increasing $[V_{o\gamma}H]$.
This phenomenon means that $V_{o\gamma}H$ acts mainly as a recombination center.
In construct, a defect acts mainly as a release center if the damage
increases with its concentration.
In Table.~\ref{tab:initial0}, we list different roles played by the six defects.
In Fig.~\ref{fig:VogH0}, we also notice that
the less the recombination center the smaller the enhancement factor.
Thus, one may expect the appearance of
an RLDRS effect
for low enough [$V_{o\gamma}H$].
However, as can be seen in Fig. \ref{fig:VogH0}, no matter how we try to tune down the concentration,
the tendency of ELDRS never changes. 

\begin{figure}[!t]
  \centering
  \includegraphics[width=0.7\linewidth]{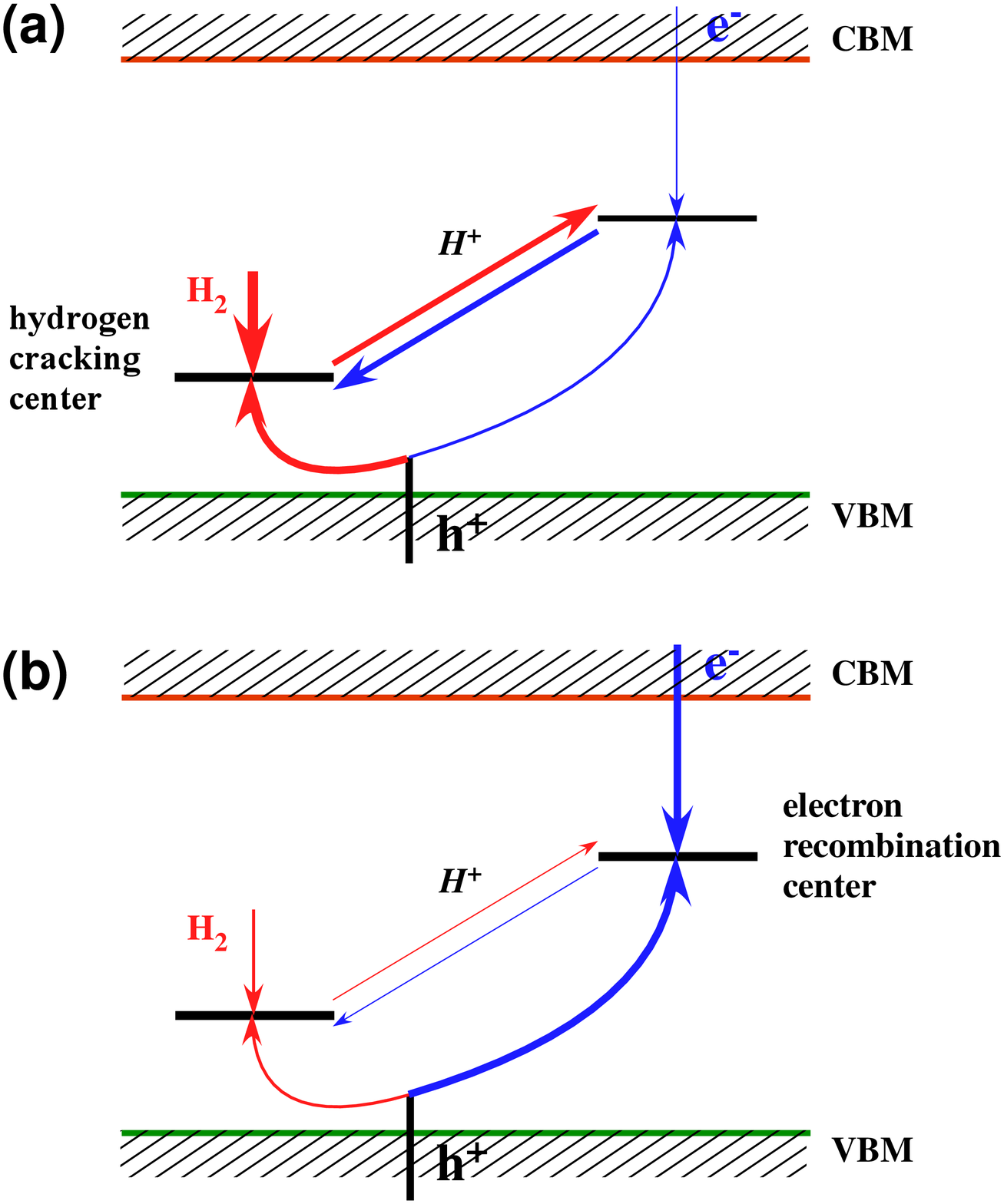}\\
  \caption{Hydrogen cracking and electron recombination reactions and their competitions
  in a-SiO$_2$ for (a) ELDRS and (b) RLDRS effects.
  The line widths stand for the difference in defect concentrations and reaction strengths.
    }\label{fig:mechanism}
\end{figure}

\emph{Model analysis.}--
Since the transport properties in a-SiO$_2$
are less influenced by the concentration of defects, 
we can just take into consideration the reaction part of the model.
Moreover, since the shallow defects 
 mainly determine the hopping transport and exist in bulk (see table \ref{tab:initial0}),
we can just consider the deep defects 
concentrating on the interface. 
From the reaction rates calculated from first principle theory \cite{rowsey2012quantitative}, we know
that $V_{o\gamma}$ and $V_{o\gamma}H$ are similar to capture a hole.
However, after capturing a hole, the former is easier to release a proton while the latter is easier
to annihilate an electron.
Therefore, the important reactions that are sensitive to dose rate
and defect concentrations are\\
\text{Hydrogen Cracking Reaction:}
\begin{subequations}
\begin{align}
V_{o\gamma}+h^+ \rightleftharpoons & ~V_{o\gamma}^+ \label{eq:reac_Vogp},\\
V_{o\gamma}^++H_2 \rightleftharpoons & ~V_{o\gamma}H+H^+, \label{eq:reac_cracking}
\end{align}\label{eq:reac_H2}
\end{subequations}
\noindent\text{and Electron Recombination Reaction:}
\begin{subequations}
\begin{align}
V_{o\gamma}H+h^+  \rightleftharpoons & ~V_{o\gamma}H^+, \label{eq:reac_recomb1} \\
V_{o\gamma}H^++e^-  \rightleftharpoons & ~V_{o\gamma}H. \label{eq:reac_recomb2}
\end{align}\label{eq:reac_e}
\end{subequations}
\noindent Here the four reactions related to $V_{o\gamma}H_2$ are 
ignored because they not much less sensitive to dose rate.
It is obvious that the trapping of holes 
is competed by the HCR 
and 
ERR, 
see Fig.~\ref{fig:mechanism} for a schematic diagram.

\emph{ELDRS mechanism.}--
There is a high symmetry between Eqs.~(\ref{eq:reac_H2}) and~(\ref{eq:reac_e}), where
$H_2$ in Eq.~(\ref{eq:reac_cracking})
corresponds to electrons in Eq.(\ref{eq:reac_recomb2}).
In the conventional case showing in Fig. \ref{fig:VogH0},
the concentration of hydrogen ($10^{13}$cm$^{-3}$)
is much higher than that of electron generated from the irradiation ($<10^{11}$cm$^{-3}$).
As a result, the HCR (Eq.~\ref{eq:reac_cracking}) dominates,
which is shown in Fig. \ref{fig:mechanism} (a). 
Supporting the dose rate increases by a times of $\lambda(>1)$, i.e.,
$\mathcal{R}\rightarrow\lambda\mathcal{R}$
(which means a shorten of time, $\tau\rightarrow\lambda^{-1}\tau$),
the generation rate of $V_{o\gamma}^+$ hence that of $H^+$ 
should increase by the same intensity.
Due to $N_{it}\propto\int_0^\tau [H^+] dt=\int_0^{\lambda^{-1}\tau} \lambda[H^+] dt$,
this change would give a same damage as that for $\mathcal{R}$.
However, the reaction described by Eq.~(\ref{eq:reac_cracking}) is reversible
(i.e., the reaction rate for the reverse and forward reactions are comparable)
and governed by the Le Chatelier's principle \cite{quilez1995students}.
As a result of the principle, the balanced generation rate of $[V_{o\gamma}^+]$ should be
some value, say $\lambda'[V_{o\gamma}^+]$, smaller than $\lambda [V_{o\gamma}^+]$.
Accordingly, the generation rate of proton becomes $\lambda'[H^+]$, which
accumulate during $\lambda^{-1}\tau$ to a value smaller than that
for dose rate $\mathcal{R}$. This phenomenon can be described as
\begin{equation}\label{ELDRS}
  N_{it}(\lambda\mathcal{R})\propto\int_0^{\tau/\lambda} \lambda'[H^+] dt\\
  <N_{it}(\mathcal{R})\propto\int_0^\tau [H^+] dt,
\end{equation}
which is just the RHDRS or ELDRS effect.
It is seen that the ELDRS effect occurs when the TID process is
dominated by HCR and the negative dose
rate dependence of ELDRS is governed by the Le Chatelier's principle.
This is completely different from the widely believed ERR
and SRH recombination.
A similar but more qualitative mechanism can be found in Ref. \cite{rowsey2012mechanisms},
which, however, also regards the forward reaction as an ELDRS mechanism.
According to Fig. \ref{fig:mechanism} (a), increasing the concentration of recombination center
only enhances the reverse reaction and hence the enhance factor;
this is reflected in Fig. \ref{fig:VogH0}.

\begin{figure}[!t]
  \centering
  \includegraphics[width=0.9\linewidth]{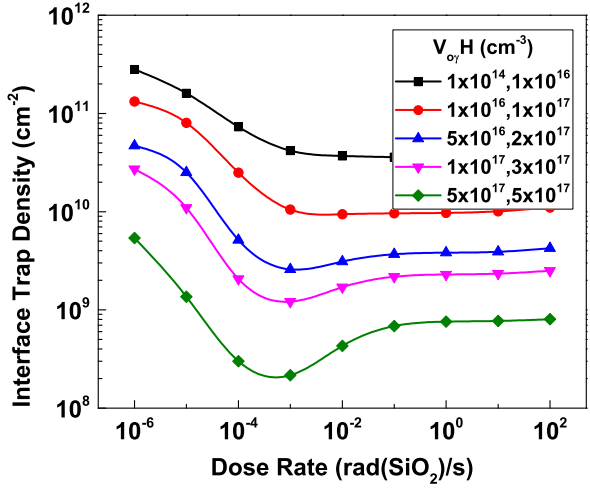}
  \caption{
  Dose rate dependence of $N_{it}$ for various concentrations of $V_{o\gamma}H$.
  The concentration of $H_2$ is $10^{10}$cm$^{-3}$.
  }\label{fig:VogH}
\end{figure}

\begin{figure}[!t]
  \centering
  \includegraphics[width=0.9\linewidth]{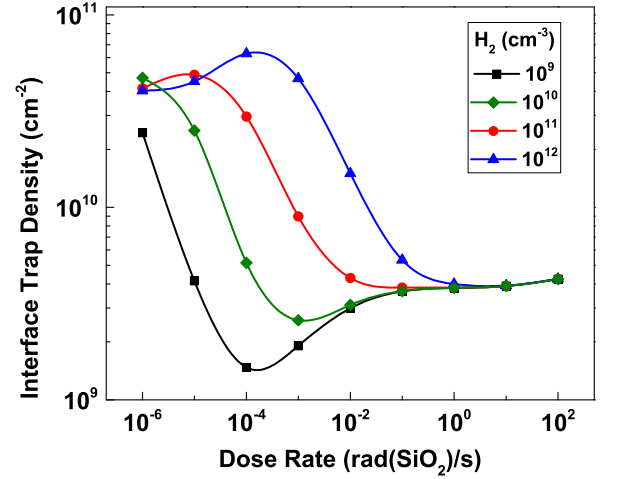}\\
  \caption{
 Dose rate dependence of N$_\textmd{it}$ for various concentrations of $H_{2}$. 
 The concentration of $V_{o\gamma}H$ is $(5\times 10^{16},2\times 10^{17})$ $\textmd{cm}^{-3}$.
  }\label{fig:H2}
\end{figure}

\emph{RLDRS with Low $H_2$.}-- 
In order to drive the system to an ERR dominated regime, 
an alternative option seems to decrease the concentration of hydrogen.
The dose-rate dependence of $N_{it}$
at $[H_2]=10^{10}\textmd{cm}^{-3}$ is plotted
in Fig.~\ref{fig:VogH}.
It is seen that the inverse correspondence between $N_{it}$ and [$V_{o\gamma}H$],
which reflects the role as a recombination center of the defect, does not change.
However, the dose rate dependence does change with increasing [$V_{o\gamma}H$].
For relatively high dose rate, the enhancement factor decreases with increasing
[$V_{o\gamma}H$] and an opposite dose rate dependence,
i.e., an RLDRS, occurs for large enough [$V_{o\gamma}H$],
see the lower three lines in Fig. \ref{fig:VogH}. 
Such a remarkable transition can be understood as following.
The decrease of [$H_2$] weakens the HCR in Eq. (\ref{eq:reac_H2})
while the increase of [$V_{o\gamma}H$] enhances the ERR in Eq. (\ref{eq:reac_e}).
As a result, the ERR becomes comparable and even dominating
for relatively high dose rates, i.e., changes from Fig. \ref{fig:mechanism} (a) to Fig. \ref{fig:mechanism} (b).
For clarity, in Fig.~\ref{fig:H2}
we plot the dose-rate dependence of $N_{it}$
for decreasing $[H_2]$ at a high [$V_{o\gamma}H$].
An ELDRS-RLDRS transition is clearly seen for relatively high dose rate,
which stems from a same reason as above.
Thus, decreasing the concentration ratio of $H_2$ and
recombination center
seems an efficient way to induce the RLDRS effect.
Using this effect, the TID damage
of BJT at a low dose rate can be evaluated at a relatively high dose rate
with a dramatically shortened time cost ($\tau\rightarrow\tau/\lambda$,
$\lambda$ is several order of magnitudes).

It is surprising that, in most known literatures only the ELDRS effect was reported. 
The above analysis shows that such a strong asymmetry maybe stem from
the fact of that the concentration of $H_2$ in a-SiO$_2$ is always much higher
than that of electron induced by irradiation.
However, we find that,
in an early experiment by Pease et al \cite{pease2008effects},
there is a trench near 1rad/s which maybe regarded as an RLDRS effect.
The trench is only observable for samples exposed to air
and disappears for 
any higher $H_2$ concentrations,
which coincides exactly with our interpretations.
Also the decrease of $N_{it}$ with decreasing [$H_2$] is the same as that shown in Fig. \ref{fig:H2}. 

\emph{RLDRS mechanism.}--
As shown in Fig. \ref{fig:mechanism}(b),
the ERR becomes dominating in presence of RLDRS.
Due to the SRH recombination theory, the amount of consumed holes through Eq. (\ref{eq:reac_e})
reads $\int_0^{\tau} [h] R dt$,
where [h] is the hole concentration induced by irradiation and 
$R$ is the SRH recombination rate. 
With the increase of $\mathcal{R}\rightarrow\lambda \mathcal{R}$,
the hole concentration increases and the time duration decreases with a same intensity:
[h]$\rightarrow\lambda$[h] and $\tau\rightarrow\lambda^{-1} \tau$,
which will lead to a same consume as $\mathcal{R}$:
$\int_0^{\lambda^{-1}\tau} \lambda[h] R dt=\int_0^{\tau} [h] R dt$.
However, the recombination rate decreases with increasing dose rate. 
This is because $R  \propto \textmd{e} ^{-(E_F^e-E_F^h)/k_BT},$
where $E_F^e$ and $E_F^h$ are the quasi-Fermi energies of electrons and holes respectively \cite{shockley1952statistics}.
The higher the dose rate the bigger the difference
between the quasi-Fermi energies $(E_F^e-E_F^h)$, hence the smaller the recombination rate
($R\rightarrow R'<R$).
As a result, 
the holes consumed from the ERR reduces
with increasing does rate,
which leaves more holes to the HCR and hence produces larger damage.
This can be described by
\begin{equation}\label{RLDRS}
  N_{it}(\lambda\mathcal{R})\propto\int_0^{\tau/\lambda} -\lambda [h] R' dt
  >N_{it}(\mathcal{R})\propto\int_0^{\tau} -[h] R dt,
\end{equation}
which is just the EHDRS or RLDRS effect.
It is seen that, the RLDRS effect occurs when the TID process
is dominated by ERR
and the abnormal positive dose-rate dependence of RLDRS
is governed by the negative dose rate dependence of SRH recombination rate.

\begin{figure}[!t]
  \centering
  \includegraphics[width=0.9\linewidth]{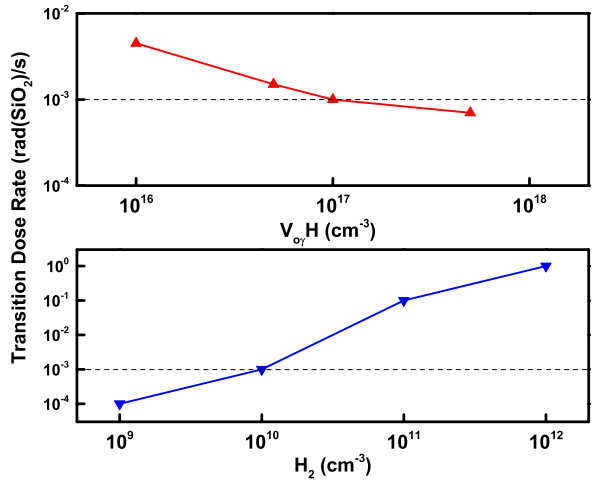}
  \caption{
 Transition dose rate as a function of the concentration of (a) $V_{o\gamma}H$ and (b) H$_2$,
 extracting from Figs. \ref{fig:VogH} and \ref{fig:H2}, respectively.
  }\label{fig:control}
\end{figure}

\emph{RLDRS-ELDRS transition.}--
It is noted that, in both Figs. \ref{fig:VogH} and \ref{fig:H2}
the RLDRS occurs only for relatively high dose rates
and evolves to the ELDRS again for lower dose rates.
This is because the relative concentration of electron becomes smaller than
that of the $H_2$ and the HCR dominates again.
At the transition dose rate, the damage is minimum.
The transition dose rate stems from the competition between the HCR and
the ERR
hence can be accurately controlled by tuning the concentration ratio of hydrogen and recombination centers.
This is clearly demonstrated in Fig.~\ref{fig:control}. 
The transition dose rate increases or decreases with increasing
concentration of hydrogen or recombination center,
reflecting the different role played by them.
Such a tunability paves a way for application
in different environments.
For the application in space, where the typical dose rate is $10^{-3}$rad(SiO$_2$)/s $\sim10^{-2}$rad(SiO$_2$)/s,
a [H$_2$] below $10^{10}\textmd{cm}^{-3}$ 
and [$V_{o\gamma}H$] higher than $(5\times 10^{16},2\times 10^{17})$$\textmd{cm}^{-3}$ is enough.
For other applications with even low dose rate,
lower [$H_2$] and (or) higher [$V_{o\gamma}H$] is required 
to assure the occuring of RLDRS.

\emph{Conclusion}.--
In summary,
we have shown that in conventional cases only the ELDRS effect occurs
because the concentration of hydrogen is much higher 
and 
the hydrogen cracking reaction dominates (see, Eq. \ref{eq:reac_H2} and Fig. \ref{fig:mechanism} (a)).
The negative dose-rate dependence of ELDRS stems from the decrease of
the generation rate of proton with increasing dose rate (see, Eq. \ref{ELDRS}),
which is a direct result of the Le Chatelier's principle.
More importantly, we have proposed that
an RLDRS effect can occur
by decreasing the concentration ratio between hydrogen and recombination centers hence
making the electron recombination reaction dominating (see, Eq. \ref{eq:reac_e} and Fig. \ref{fig:mechanism} (b)).
The abnormal positive dose-rate dependence of RLDRS stems from the
decrease of the SRH recombination rate with increasing dose
rate (see, Eq. \ref{RLDRS}).
We have also found an RLDRS-ELDRS transition for decreasing dose rate
and its controllability by the concentration ratio.
The proposed RLDRS effect paves a way for a credible acceleration test
of the TID damage of BJT at certain low dose rates
while the tunability provides a strategy for applications in different dose rate environments.

We have to emphasize that, the above proposal and mechanisms rely only on the competition
between hydrogen cracking and electron recombination reactions,
while the details of the concentration profile and other parameters
such as reaction rates are not important. 
We encourage experiments on the proposed effect and mechanisms.

\emph{Acknowledgements}.--
This work was supported by the Science Challeng Project (Grant No. TZ2016003-1)
and the National Natural Science Foundation of China (Grant No. 11404300).
XLF, ZBW, and YL contribute equally to this work.


%

\end{document}